\noindent

\vglue 2cm \font\bigbold=cmr10 scaled\magstep3
 { \bigbold Canonical  bases for real representations of
Clifford algebras}

\vskip 0.5cm

\vskip 16pt \item {} \hskip 3.5cm
 {\font\big=cmr10 scaled\magstep2 {\big A. H. Bilge $^a$},
 {\font\big=cmr10 scaled\magstep2 {\big \c S. Ko\c cak $^b$},
 {\font\big=cmr10 scaled\magstep2 {\big S. U\u guz $^{c,}$
}
{\font\smallnine=cmr9 scaled\magstep0{\smallnine\footnote
*{Corresponding Author. Tel.: +90-212-2856795; fax:+90-212-2856386
\item {}{\font\smallnine=cmti9 scaled\magstep0{\smallnine E-mail
Addresses:}} bilge@itu.edu.tr (A.H. Bilge), skocak@anadolu.edu.tr
(\c S. Ko\c cak), uguzs@itu.edu.tr (S. U\u guz). }}

\vskip0.5cm
\leftline{$^a${\it Department of
Mathematics, Istanbul Technical University, Maslak, Istanbul,
 Turkey} and }
\leftline{\quad
{\it TUBITAK, Feza Gursey Institute, \c Cengelkoy,
Istanbul, Turkey}}
\leftline{$^b${\it Department of Mathematics, Anadolu University,
Eskisehir, Turkey}}
\leftline{$^c${\it Department of
Mathematics, Istanbul Technical University, Maslak, Istanbul
 Turkey} }

\vskip 1cm
 \hrule   width16.5cm height0.7pt
\vskip 1cm

\noindent {\bf Abstract}
\vskip 0.5cm

 The
well-known classification of the Clifford algebras $Cl(r,s)$ leads to
 canonical forms of complex and real representations which are
essentially unique by virtue of the
Wedderburn theorem.
For  $s\ge 1$  representations of $Cl(r,s)$  on $R^{2N}$
 are obtained from representations on $R^N$ by adding two
new generators while
in passing from a representation of $Cl(p,0)$  on $R^N$ to a
representation of $Cl(r,0)$ on
 $R^{2N}$
the number of  generators that can be added is either $1$, $2$ or
$4$, according as the Clifford algebra represented on $R^N$ is of
real, complex or quaternionic type. We have expressed canonical
forms of these representations in terms of the complex and
quaternionic structures in the half dimension and we obtained
algorithms for transforming any given representation of $Cl(r,s)$
to a canonical form. Our algorithm for the transformation of the
representations of $Cl(8d+c,0)$, $c\le 7$  to canonical forms is
based on finding an abelian subalgebra of $Cl(8d+c,0)$ and its
invariant subspace.  Computer programs for determining explicitly
the change of basis matrix for the transformation to canonical
forms are given for lower dimensions. The construction of the
change of basis matrices uniquely up to the commutant provides a
constructive proof of the uniqueness properties of the
representations and may have applications in computer graphics and
robotics.

\vskip 1cm

\hrule   width16.5cm height0.5pt \vskip 0.7cm

\vskip  1cm \baselineskip 16pt
\noindent
{\bf 1. INTRODUCTION}

The classification and representation of  Clifford
algebras is well-known:
Any real Clifford algebra  $Cl(r,s)$ is
isomorphic to one of the matrix algebras $K(2^n)\cong End(K^{2^n})$ or
$K(2^n)\oplus
K(2^n)$,
 where $K$ is either
 the reals $R$, the complex numbers $C$ or the quaternions $H$ [1].
The proof of the classification theorem is
constructive
 and  can be used directly to build
real representations in the case $s\ge 1$
while the problem is more tricky for  $Cl(r,0)$.
The construction of complex representations is also straightforward.

The structure of the representations is
based on  the
Wedderburn theorem which states that the matrix algebras $K(N)$
have a unique representation on the vector space $K^N$ [2]. It follows
that the Clifford algebra
$Cl(r,s)$ has exactly one or two representations on $K^{2^n}$,
according as it
is isomorphic to $K(2^n)$ or to $K(2^n)\oplus K(2^n)$.  The
uniqueness
 of the representation over $K$
means that any two  representation can be transformed to each other
by conjugation with a unique matrix   $P$ with entries in $K$.

The main result of our paper, given in Section 4,  is the
construction of an orthonormal basis for the representation space
with respect to which the matrices in any given representation
have desired canonical forms. We use here the word ``canonical''
for representations expressible as homogeneous tensor products of
the standard Pauli matrices (Eq.2.6). Computer programs leading to
the corresponding change of basis matrix $P$ in lower dimensions
are presented  in Appendix A.

In Section 2 we give a concise overview of the classification and
representation of Clifford algebras, as the proof of the
classification theorem is the key in understanding the construction
of canonical representations. As an immediate corollary of the
classification theorem, we give the formula for the algebraic type
of a real representation of $Cl(r,s)$ in terms of $r$ and $s$
(Proposition 2.8),
previously obtained using the
representation theory
of finite groups [3,4].

Note that if a Clifford algebra $Cl(r,s)$ is isomorphic to a  matrix
 algebra over complex or quaternionic numbers
and we use a real
 representation, then the matrices in the complex or quatrernionic
 subalgebras will commute with all matrices of the representation. Such matrices are called the ``commutant'' [4].

 In Section 3 we obtain the relations
between maximal Clifford algebras that can be represented on
$R^{2N}$  and the structure of the  commutant in the half dimension.
The classification theorem states that the generators of
$Cl(p,q)$  can be expressed as a tensor product of the generators
of $Cl(r,s)$  with $p+q=r+s+2$  and the generators of $Cl(1,1)$,
$Cl(0,2)$ or $Cl(2,0)$  (Lemmas 2.3 and 2.4).
As $Cl(1,1) $ and $Cl(0,2)$ have 2-dimensional real representations,
the representations of  Clifford algebras $Cl(r,s)$ with
 $s\ge 1$ follow directly from the classification theorem.
The difficulty with  the representations of $Cl(r,0)$  comes from the fact that
irreducible representations of $Cl(2,0)$ are 4-dimensional, hence
the addition of
$2$ or more generators as one doubles the dimension is a non-trivial
 problem.
In the representations of $Cl(r,0)$ the number of generators to be added as we
 double the dimension  is $1$, $2$ or $4$, depending on   whether the
representation in the half dimension
 is real, complex or quaternionic.  It turns out that the
 possibility of adding more than one generator  is
due to existence of this  nontrivial commutant.
On the other hand, the existence of a nontrivial commutant is tied
to the structure of the maximal Clifford algebra $Cl(r,s)$ that
can be represented in the half dimension. Namely, the extendibility of
a representation of  $Cl(8d,0)$ to a representation of $Cl(8d,1)$ at
 the same dimension leads to the complex structure for the
 representations of $Cl(8d+1,0)$.  Similarly the extendibility of
 $Cl(8d+1,0)$ to $Cl(8d+1,2)$ leads to the quaternionic structure for
 $Cl(8d+3,0)$.
The interrelations between these structures are displayed in Table
2.

In Section 4 we study the problem of transforming a given real
irreducible representation of a Clifford algebra to a  canonical form.
 As noted above, the Wedderburn theorem implies
that if $A_i$'s and $\tilde{A}_i$'s are arbitrary representations
of $Cl(r,s)$, there is a matrix $P$ unique up to the commutant,
such that $A_iP=P\tilde{A}_i$ for $i=1,\dots , r+s$, but the
determination of such a matrix $P$ is  nontrivial especially for
the representations of $Cl(r,0)$.  We describe below  the
difficulties involved and outline our  solution.

Recall that as $Cl(1,1)$ and $Cl(0,2)$ have 2-dimensional
representations, one can easily construct representations on $2N$
dimensions with tensor products using representations on $N$
dimensions. The converse problem is to ``recognize'' the
generators of $Cl(1,1)$ or $Cl(0,2)$ and express the remaining
elements as tensor products.  This is easy because any two
anti-commuting elements in the representation with squares $\pm I$
as appropriate (Lemma 4.1),  can  be put to canonical forms and
any matrix anti-commuting  with the two anti-commuting elements
has a block-diagonal structure (Proposition 4.2).

For representations of $Cl(r,0)$ it is easy to put one generator
to a canonical form, but this does not lead to
block-diagonalization. Hence the converse problem is nontrivial
even  for the case when only a single generator is added in
passing to the double dimension. For transforming the generators
of $Cl(r,0)$ to canonical forms we shall use an algorithm
mimicking the situation for  $Cl(3,0)$. If  $A_1$, $A_2$ and
$A_3=A_1A_2$ belong to a representation of $Cl(3,0)$ on $R^4$,
they can be transformed to ``canonical'' quaternionic structures
by choosing a basis $\{X,-A_1X,-A_2X,-A_1A_2X\}$ where $X$ is an
{\it arbitrary} unit vector (Proposition 4.5). This construction
cannot be used for $Cl(r,0)$ with $r\ge 3$, because the products
of the images of the generators are in general linearly
independent matrices. To apply this procedure to higher
dimensions, we find a special vector $X$ and a subalgebra $\cal A$
generated by a certain subset of the generators such that the
action of $\cal A$ on $X$  leads to the required basis.

To illustrate the procedure, consider a representation of
$Cl(6,0)\cong R(8)$ on $R^8$.  The images of the standard
generators $A_i$, $i=1,\dots, 6$ generate the matrix algebra
$R(8)$ in which the diagonal matrices constitute an 8-dimensional
maximal abelian subalgebra that we denote by $\cal D$. We aim to
express  this abelian subalgebra in terms of the generators of the
representation and in  Proposition 4.7 we show that
$\{A_1A_2A_3,A_1A_4A_5,A_2A_4A_6\}$ consisting of matrices with
squares $I$ is a generating set.
 As a set of commuting diagonalizable matrices, they are
simultaneously diagonalizable and in addition, they have a unique
common eigenvector $X$ corresponding to the eigenvalue $1$. As
$\cal D$ acts as identity on the one dimensional subspace spanned
by $X$, only three of the $A_i$'s $i=1,\dots ,6$  are independent
and the action  of the subalgebra $\cal A$ generated by
$\{A_1,A_2,A_4\}$ on $X$ generates the required basis. The
construction for higher dimensional real representations is
similar, but for complex and quaternionic representations the
vector $X$ is not unique. The uniqueness of the vector $X$ up to
the commutant leads to an alternative proof of the uniqueness of
the representations up to the commutant.

With continuing interest in the relations to group representations
[5], representations of Clifford algebras are now finding
applications in the  field of robotics and computer graphics
[6-8]. In these approaches, the motions of rigid bodies in
3-space are modelled with Clifford algebras of various types
related to quaternions.  These Clifford algebras are representable
on $R^8$ and the translation  of the data from  one coordinate
system to another is a basic problem for which the transformation
algorithms given in Appendix A are expected to be useful.

\vskip 1cm
\noindent
{\bf 2. CLASSIFICATION AND REPRESENTATION OF
CLIFFORD ALGEBRAS}

\vskip 0.2 cm

In this section we give an overview of the classification and
representation of Clifford algebras, based on the presentation in
[1].  In Section 2.1 we introduce the notation and give basic
definitions. We have also included a section on the classification
and representation of complex Clifford algebras for completeness.
In Section 2.3 we give the classification of real Clifford
algebras and we conclude with the determination of  the algebraic
type of a representation of $Cl(r,s)$ in terms of $r-s$ and $r+s$.
 Propositions 2.7 and 2.8  provide an alternative derivation of
some of the results given in [4].

\vskip 0.2cm

\noindent
{\bf 2.1 Basic definitions.}
\vskip 0.2 cm

Let $V$ be a vector space over the field $k$ and $q$ be a
quadratic form on $V$. The {\it Clifford algebra} $Cl(V,q)$
associated to $V$ and $q$ is an associative algebra with identity
$1$, generated by the vector space $V$ and by the identity,
subject to the relations $v \cdot v=-q(v) 1$ for any vector $v$ in
$V$. The map $\alpha(v)=-v$ for $v\in V$ extends to an involution
of the Clifford algebra $Cl(V,q)$ and its $\pm 1$ eigenvalues are
called respectively {\it even} and {\it odd} parts. Furthermore,
the Clifford algebra and the exterior algebra of $V$ are
isomorphic as vector spaces. The {\it order} of a Clifford algebra
element is defined as its order as an exterior algebra element.

The real
Clifford algebras associated to $V=R^{r+s}$ and to the quadratic
form $q(x)=x_1^2+\dots+x_r^2-x_{r+1}^2-\dots-x_{r+s}^2,$ is
denoted by $Cl(r,s)$. For $V=C^{n}$, as  all non-degenerate
quadratic forms over $C^n$ are equivalent, $q(z)$ is necessarily
$q(z)=z_1^2+\dots +z_n^2$. The corresponding complex Clifford
algebra is denoted by $Cl_c(n)$ [1]. If  $ \{ e_1,e_2,...,e_n\}$
is an orthonormal basis for $V$, the real  Clifford algebra
$Cl(r,s)$ is generated by the  $\{e_i\}$'s, subject to the
relations,
$$ e_i^2 =-1,\quad i=1,\dots , r\quad\quad
e_{r+i}^2 =1,\quad i=1,\dots , s, \quad\quad
e_i e_j+e_j e_i=0, \quad i\ne j.\eqno(2.1)$$
Similarly, the generators of the complex Clifford algebra $Cl_c(n)$
satisfy
$$e_i^2 =-1,\quad i=1,\dots , n,\quad\quad
  e_i e_j+e_j e_i=0, \quad i\ne j.\eqno(2.2)$$
Both $Cl(r,s)$ with  $r+s=n$ and $Cl_c(n)$ are $2^n$-dimensional vector spaces
spanned by the set
$$\{1,e_1,e_2,...,e_{r+s},e_1e_2,...,
              e_{r+s-1}e_{r+s},e_1e_2e_3,...,e_1e_2e_3...e_{r+s}\}.
\eqno(2.3)$$

If $K$ is a division algebra  containing the field $k$,
a {\it $K$-representation} of the Clifford algebra $Cl(V,q)$ on the finite
dimensional $K$-vector space $W$, is a $k$-algebra
homomorphism
$$\rho: Cl(V,q)\to Hom_K(W,W).\eqno(2.4)$$
A representation is called {\it reducible} if $W$ can be written
as a nontrivial direct sum of $\rho$ invariant subspaces.  A
representation which is not reducible is called {\it irreducible}.
It is known that every $K$-representation $\rho$ of a Clifford
algebra $Cl(V,q)$ can be decomposed into a direct sum
$\rho=\rho_1\oplus \dots \oplus \rho_m$ of irreducible
representations.

Two  representations $\rho_j:Cl(V,q)\to Hom_K(W_j,W_j)$ for $j=1,2$ are said to
be {\it equivalent} if there exists a $K$-linear isomorphism $F:W_1\to W_2$
such that $F\circ \rho_1(\varphi)\circ F^{-1}=\rho_2(\varphi)$ for all
$\varphi\in Cl(V,q)$.
In particular, for  $W_1=W_2$ if two representations are equivalent,
$\rho_2(\phi)$ is obtained from $\rho_1(\phi)$ for each $\phi$ in $Cl(V,q)$, by
conjugation with the same matrix.

The algebra of linear endomorphisms of $K^N$, $End(K^N)$,  is
denoted by $K(N)$ It is known that these matrix algebras are
simple and have a unique representation up to equivalence [2].
This result known as the Wedderburn theorem also determines  the
structure of the representations of Clifford algebras.

\proclaim  Proposition 2.1. (Wedderburn theorem) Let $K=R,C$
or $H$ and consider the ring $K(N)$ of $N\times N$ matrices as an
algebra over $R$. Then the natural representation $\rho$ of $K(N)$
on the vector space $K^N$
 is, up to equivalence, the only irreducible representation of $K(N)$.
The algebra $K(N)\oplus K(N)$ has exactly two irreducible
representations given by
$$\rho_1(\phi_1,\phi_2)=\rho(\phi_1),\quad
  \rho_2(\phi_1,\phi_2)=\rho(\phi_2),   \eqno(2.5) $$
 where $\rho$ is the natural representation.

We give below certain isomorphisms that are needed in proofs.
 \proclaim Proposition 2.2. (Proposition 4.2 in [1])There are isomorphisms
$$\eqalignno{
R(n)\otimes_ R R(m)  \cong &R(nm),\quad R(n)\otimes_R K    \cong
K(n), \quad \quad K=C,H,\cr
 C   \otimes_R C     \cong & C\oplus
C,\quad C   \otimes_R H    \cong   C(2),\quad H \otimes_R H \cong
R(4),\cr
(K(n)\oplus K(n))\otimes_K K(m)\cong & K(nm)\oplus K(nm),
\quad \quad K=R,C \cr }$$

\noindent Finally we present  our notation. In 2-dimensions the
standard Pauli matrices are denoted as
$$\sigma   =\pmatrix{1&0\cr 0&-1\cr},\quad
  \epsilon =\pmatrix{0&1\cr -1&0\cr},\quad
  \tau     =\pmatrix{0&1\cr  1&0\cr},\quad\eqno(2.6)$$
and their multiplication rules are
$$\sigma\tau=\epsilon,\quad \sigma\epsilon=\tau,\quad
   \epsilon\tau=\sigma.\eqno(2.7)$$
The tensor products are expressed as
$$\sigma   \otimes a=\pmatrix{a&0\cr 0&-a\cr},\quad
  \epsilon \otimes b=\pmatrix{0&b\cr -b&0\cr},\quad
  \tau     \otimes c=\pmatrix{0&c\cr  c&0\cr}.\quad\eqno(2.8)$$

For simplicity of notation, in tensor products, identity matrices
of any size will be denoted by $1$ unless  the distinction is
important. Also by abuse of language, the images of the generators
of a Clifford algebra under a representation are called the {\it
generators of the representation}.

\vskip 0.2 cm

\noindent {\bf 2.2. Complex Clifford  algebras.}
\vskip 0.2 cm

We give the
classification and representation of complex Clifford algebras.

\proclaim Lemma 2.3.  For all $n\ge 0$, there are isomorphisms
$$Cl_c(n+2)\cong
Cl_c(2)          \otimes_C           Cl_c(n),\eqno(2.9a)$$
$$Cl_c(2n)\cong
C(2^n),\quad\quad Cl_c(2n+1)\cong C(2^n)\oplus
C(2^n).\eqno(2.9b)$$

\noindent {\bf Proof.} Let $e_j$, $j=1,\dots ,n$ be the generators
of $Cl_c(n)$, and let $e'_1$ and $e'_2$ be the generators of
$Cl_c(2)$.  Then
$$\tilde{e}_j    = ie'_1e'_2\otimes e_j,\quad j=1,\dots n,\quad
  \tilde{e}_{n+1}= e'_1\otimes 1,\quad
  \tilde{e}_{n+2}= e'_2\otimes 1,\quad
  \eqno(2.10)$$
is a set of generators for $Cl_c(n+2)$, hence Eq.(2.9a) is proved.
Eq.(2.9b) follows from Eq.(2.9a) and from the fact that
$Cl_c(1)\cong C\oplus C$ and $Cl_c(2)\cong C(2)$, using  the
isomorphisms in Proposition 2.2. \hfill$\bullet$

As there is a 2-periodicity, the construction of the representations is
straightforward.
Starting with a representation  of  $Cl_c(3)$ on $C^2 $ as
$$
  \rho(e_1)=i\pmatrix{0&1\cr 1& 0\cr},\quad
  \rho(e_2)=\pmatrix{0&1\cr -1& 0\cr},\quad
  \rho(e_3)=i\pmatrix{1&0\cr 0& -1\cr},\quad
\eqno(2.11)$$
and given any
representation of $Cl(n)$ on $C^N$, an irreducible representation of $Cl_c(n+2)$
can be obtained by replacing the generators with their representations in the
proof of Lemma 2.3.  Namely, if
$\rho(e_j)=a_j,\quad j=1,\dots ,n$
is a representation on $R^N$, then
$$\rho(\tilde{e}_j)    =i\sigma \otimes a_j  =
\pmatrix{a_j & 0 \cr  0 & -a_j\cr},\quad
  \rho(\tilde{e}_{n+1})=\epsilon\otimes 1 = \pmatrix{0&1\cr -1&0 \cr},\quad
  \rho(\tilde{e}_{n+2})=i\tau\otimes 1  =i\pmatrix{0&1\cr 1&0 \cr},\quad
\eqno(2.12)$$
gives a representation on $R^{2N}$.

If $Cl(r,s)$ is a Clifford algebra isomorphic to $C(2^n)$, it will
have an irreducible representation on $R^{2^{n+1}}$.  These
representations can be obtained from a complex representation on $C^{2^{n}}$
once the complex structure is known.
This provides an alternative method for the
construction of representations of $Cl(8d+1,0)$, because the product
of
the generators is
an odd Clifford algebra element which is central, hence it is the complex
structure $J$ (see Proposition 4.4).

\vskip 0.2 cm

\noindent {\bf 2.3.  Real Clifford  algebras.}
\vskip 0.2 cm

The crucial step in the classification of real Clifford algebras
is the isomorphism theorem below.

\proclaim Lemma 2.4. (Theorem
4.1 in [1]) There are isomorphisms
$$Cl(1,1)\otimes Cl(r,s)\cong Cl(r+1,s+1),\eqno(2.13a)$$
$$Cl(0,2)\otimes Cl(r,s)\cong Cl(s,r+2
),  \eqno(2.13b)$$
$$Cl(2,0)\otimes Cl(r,s)\cong Cl(s+2,r),  \eqno(2.13c)$$
for all $r$, $s\ge 0$.

\noindent {\bf Proof.} The generators of $Cl(r,s)$ with squares
$-1$ are denoted by $a_i$ and the ones with square $+1$ by $b_i$.
Let $(a_1', b_1')$, $(a_1', a_2')$, $(b_1', b_2')$ be the
generators of $Cl(1,1)$, $Cl(2,0)$ and $Cl(0,2)$ respectively.
Note that as $a_i^2=-1$, $b_i^2=1$ and as they form an
anti-commuting set, $(a_1'b_1')^2=1$ while,
$(a_1'a_2')^2=(b_1'b_2')^2=-1$. Thus there will be a twisting
whenever a Clifford algebra element of $Cl(r,s)$ is tensored with
an element of $Cl(2,0)$ or $Cl(0,2)$. Thus
$$\eqalignno{
(a_1'b_1'\otimes a_i)^2=&-1,\quad\quad  (a_1'b_1'\otimes b_i)^2=1,\cr
(a_1'a_2'\otimes a_i)^2=&1,\quad\quad   (a_1'a_2'\otimes b_i)^2=-1,\cr
(b_1'b_2'\otimes a_i)^2=&1,\quad\quad   (b_1'b_2'\otimes b_i)^2=-1.&(2.14)\cr
}$$
Hence the generators of $Cl(r+1,s+1)$ can be obtained easily as
$$\eqalignno{
\tilde{a}_i=&a_1'b_1'\otimes a_i,\quad i=1,\dots ,r,\quad\quad
                                         \tilde{a}_{r+1}=a_1'\otimes 1\cr
\tilde{b}_i=&a_1'b_1'\otimes b_i,\quad i=1,\dots ,s,\quad\quad
                                         \tilde{b}_{s+1}=b_1'\otimes 1.
&(2.15)\cr
}$$
A similar construction works for $Cl(s,r+2)$ with a twisting.  The standard
generators are obtained as
$$\eqalignno{
\tilde{a}_i=&b_1'b_2'\otimes b_i,\quad i=1,\dots ,s,\quad\quad
                                              \cr
\tilde{b}_i=&b_1'b_2'\otimes a_i,\quad i=1,\dots ,r,\quad\quad
                \tilde{b}_{r+1}=b_1'\otimes 1, \quad\quad
\tilde{b}_{r+2}=b_2'\otimes 1
&(2.16)\cr
}$$
Finally the generators of $Cl(s+2,r)$ are also obtained with a twist as
$$\eqalignno{
\tilde{a}_i=&a_1'a_2'\otimes b_i,\quad i=1,\dots ,s,\quad\quad
                  \tilde{a}_{s+1}=a_1'\otimes 1, \quad
\tilde{a}_{s+2}=a_2'\otimes 1
                                              \cr
\tilde{b}_i=&b_1'b_2'\otimes a_i,\quad i=1,\dots ,r,\quad\quad
&(2.17)\cr
}$$
and the proof is completed.
\hfill$\bullet$

The proofs of the isomorphisms (2.13a) and (2.13b) lead directly
to the construction of real representations of $Cl(r,s)$ for $s\ne
0$. However the construction of the real representations of
$Cl(r,0)$ is not a direct consequence of the  proof of (2.13c),
because the generators of $Cl(2,0)$ are represented by $4\times 4$
matrices.

The Clifford algebras $Cl(n,0)$ and $Cl(0,n)$ for $n\le 8$ as
given in [1, Section 1, Table 1] in proving Proposition 2.6, and
can be read off from Eqs 2.19a-h we do not list them here.

Iterating the isomorphisms in Lemma 2.4, we can obtain the
``periodicity isomorphisms" (Theorem 4.3, in [1]) as follows.

\proclaim Proposition 2.5.
There are isomorphisms
$$\eqalignno{
Cl(0,n+8) \cong& Cl(0,n)\otimes Cl(0,8),&(2.18a)\cr
Cl(n+8,0) \cong& Cl(n,0)\otimes Cl(8,0),&(2.18b)\cr
Cl(r+8d,s)\cong& Cl(r,s+8d)\cong Cl(r,s)\otimes R(2^{4d})\quad r,s\le 7,&(2.18c)\cr
Cl(n+r,n+s)\cong& Cl(r,s)\otimes R(2^{n}),\quad r,s\le 7.&(2.18d)\cr
}$$

\noindent
We can then obtain the classification of $Cl(r,s)$ as

\proclaim Proposition 2.6. There are isomorphisms
$$\matrix{
Cl(n,n)  \cong    R(2^n),                  & &(2.19a)\cr \cr
\quad\quad\quad Cl(n,n+1)\cong  R(2^n)\oplus R(2^n),       &
Cl(n+1,n) \cong C(2^n)                         & (2.19b)\cr \cr
Cl(n,n+2)\cong  R(2^{n+1}),                 & Cl(n+2,n)\cong
H(2^n)                        &     (2.19c)\cr\cr Cl(n,n+3)\cong
C(2^{n+1}),                 & \quad\quad\quad\quad Cl(n+3,n)\cong
H(2^n) \oplus H(2^n)          &   (2.19d)\cr\cr Cl(n,n+4)\cong
H(2^{n+1}),                 & \quad Cl(n+4,n)\cong H(2^{n+1}) &
(2.19e)\cr\cr \quad\quad\quad\quad\quad Cl(n,n+5)\cong
H(2^{n+1})\oplus H(2^{n+1})&\quad Cl(n+5,n)\cong C(2^{n+2}) ,
&(2.19f)\cr\cr Cl(n,n+6)\cong  H(2^{n+2}),                 &\quad
Cl(n+6,n)\cong R(2^{n+3})                    &  (2.19g)\cr\cr
Cl(n,n+7)\cong C(2^{n+3}),                 &
\quad\quad\quad\quad\quad\quad Cl(n+7,n)\cong R(2^{n+3}) \oplus
R(2^{n+3})  & (2.19h)\cr\cr }$$

\noindent {\bf Proof.} From Eq. (2.13a) of Lemma 2.4,
$Cl(n,n)\cong Cl(n-1,n-1)\otimes Cl(1,1)$. Then (2.19a) can be
proved by induction using the isomorphism $Cl(1,1)\cong R(2)$. The
proofs of (2.19b-h) are similar.\hfill$\bullet$

\noindent
The isomorphisms (2.19a-h) can be rearranged   in the format below.
\proclaim
 Proposition 2.7. The Clifford algebras $Cl(r,s)$ are isomorphic to either
of the matrix algebras $R(2^n)$, $R(2^n)\oplus R(2^n)$, $C(2^n)$, $H(2^n)$ or
$H(2^n)\oplus H(2^n)$ according to the values of $r$ and $s$ as given
below.
$$\eqalignno{
R(2^n):\quad &   s+r=2n,        \quad s-r=0,2 \quad (mod\ 8)      &
(2.20a)\cr R(2^n)\oplus R(2^n):\quad & s+r=2n+1 ,\quad   s-r=1
\quad (mod\ 8) &      (2.20b)\cr
C(2^n):\quad & s+r=2n+1,\quad
s-r=3,7 \quad (mod\ 8) &      (2.20c)\cr
H(2^n):\quad & s+r=2n+2,
\quad s-r=4,6 \quad (mod\ 8) & (2.20d)\cr
H(2^n)\oplus H(2^n):\quad
& s+r=2n+3, \quad  s-r=5 \quad (mod\ 8) & (2.20e)\cr
}$$

Recall
that a Clifford algebra isomorphic to $K(n)$ or $K(n)\oplus K(n)$
is called respectively of {\it real}, {\it complex} or {\it
quaternionic type}, according as $K=R$, $K=C$ or $K=H$. The
discussion above, together with Proposition 2.5, leads to the
classification
 of the type of the representation according to the values of $s-r$ (mod 8).  A
 proof of this theorem is given in [3,4], using finite group
 representations.  The result below  follows immediately from the
classification  theorem.
\proclaim Proposition 2.8. The Clifford  algebras $Cl(r,s)$
are of real, complex or quaternionic types respectively  according
as $s-r=0,1,2$, $s-r=3,7$ or $s-r=4,5,6$ (mod 8).

 \vskip 0.2cm
\noindent {\bf Remark 2.9} The maximal number of linearly
independent vector fields on the sphere $S^{N-1}$ is known as the
{\it Radon-Hurwitz number}  $k(N)$  computed as follows. If $N=
(2a+1) 2^{4d+c}$, $ c=0,1,2,3 $, then $k(N)=8d+2^c -1$. By
Proposition 7.1 in [1], representations of $Cl(r,0)$ on $R^N$ give
linearly independent vector fields on $S^{N-1}$, hence the
irreducible representations of $Cl(k(N),0)$ are $N$-dimensional.

\vskip 1cm
\noindent {\bf 3. CANONICAL FORMS OF REPRESENTATION }

\vskip 0.2 cm
 In this section we obtain canonical  expressions for the
 representations of $Cl(r,s)$, in the sense of homogeneous tensor
 products  of two dimensional representations of the standard
 generators.
 We prefer to work with a form where
the generators coming from the half dimension are represented by block
diagonal matrices while it would as well be possible to
represent them with off diagonal blocks.

    The constructions for $s\ge 1$ given
 in Section 3.2 are
 straightforward, while the constructions for
$Cl(r,0)$ are nontrivial and closely related to the  commutants of
 representations in half dimensions.

\vskip 0.2 cm
\noindent
{\bf 3.1. Preliminaries.}
\vskip 0.2 cm

A representation of the Clifford algebra $Cl(r,s)$ on $R^N$
determines an $r+s$-dimensional subspace in $End(R^N)\cong  R(N)$.
The images of the standard generators are linear transformations
with square $\pm I$.
 We first note that without loss of generality one can represent the standard
generators with symmetric or skew-symmetric matrices [1,
Proposition 5.16].

\proclaim Proposition 3.1.  Let  $A_i$, $i=1,\dots,r+s$ be an
anti-commuting set of endomorphisms of $R^N$  satisfying
$A_i^2=\epsilon I $ where $\epsilon=\pm 1$.  There is an  inner
product on $R^N$ with respect to which $A_i$'s are skew symmetric
or symmetric, according as $\epsilon=-1$ or $\epsilon=1$.

\noindent
{\bf Proof.}  Let $(X,Y)$ denote the
standard inner product on $R^N$ and define a new inner product
$$\eqalignno{
\langle X,Y\rangle=&(X,Y)+\sum_i (A_iX,A_iY)
+\sum_{i<j} (A_iA_jX,A_iA_jY)
+\sum_{i<j<k} (A_iA_jA_kX,A_iA_jA_kY)
+\dots             \cr
&\quad +\sum_{i_1< \dots <i_{r+s}}
(A_{i_1}\dots A_{i_{r+s}}X,A_{i_1}\dots A_{i_{r+s}}Y).
}$$
It can be checked that if $A_i^2=\epsilon I$, then $\langle
A_iX,Y\rangle-\epsilon \langle X, A_iY\rangle=0$, hence $A_i$ is
symmetric or skew-symmetric and the proposition is proved.
                       \hfill$\bullet$

  Let $S^{(N)}$ be the set of matrices in
$R(N)$ with minimal polynomial $A^2+\lambda I=0$, where $\lambda$
can be positive, negative or zero. If $\lambda $ is positive, as
complex eigenvalues occur in conjugate pairs, the eigenspaces of
$A$  have equal dimension and $A$ is trace zero. However, for
$\lambda$ zero or negative (or for complex representations) this
is no longer true. Nevertheless we  show that if  at least two
such matrices  lie in the same linear subspace their eigenspaces
have the same dimension and they are trace zero.

\proclaim Lemma 3.2. Let $A$ and $B$ be $2n\times 2n$ matrices
satisfying $A^2+\lambda  I=0$, $B^2+\mu  I=0$, $\lambda \ne 0$,
$\mu\ne 0$,  and $AB+BA=0$. Then the eigenspaces of $A$ and $B$
have equal dimension.

\noindent {\bf Proof.} We may assume that $A$ is in
Jordan canonical form over $C$, i.e $A=\sqrt{-\lambda}
\pmatrix{I_p&0\cr 0& -I_q\cr}$
 with $p+q=2n$, where $ p$ and $q$
may be unequal. Multiple by $B$ the equation $AB+BA=0$ and take
its trace to get $0= 2tr(AB^2) = -2\mu tr(A)$. Here we used the
fact that $tr(BAB) = tr(AB^2)$. This shows that $tr(A)=0$ and
hence, $p=q$. Similar argument on $B$ shows Lemma 3.2.

An important implication of this result is the following.

\proclaim Corollary 3.3.  Let $A_i$ $i=1,\dots ,n$, $n\ge 2$ be a
set of anti-commuting matrices with $A_i^2=\pm I$.  Then
$tr(A_i)=0$ for $i=1,\dots ,n$.

The Clifford algebras that can be represented at the same
dimension are shown in Table 1, which is  the
classification table in [1], where we joined the cells to display the
 Clifford algebras
that are represented at the same dimension. Given an irreducible
representation of  $Cl (r,s)$ on $R^N$, if no new generator can be
added without increasing the dimension, the representation is
called {\it maximal}. For example the representations of $Cl
(7,0)$, $Cl (3,4)$ and $Cl(4,1)$ on $R^8$ are  all maximal. If
$Cl(p,q)$ is not maximal, then the set of generators that can be
added to get a representation on the same dimension is called {\it
complementary generators}.  Note that the set of complementary
generators is not unique. For example starting with a
representation of $Cl(2,0)$ on $R^4$,  one can either add one
complementary generator with negative square to get $Cl(3,0)$, or
three  complementary generators with positive squares to get
$Cl(2,3)$.

In the construction of canonical representations, if $r$ and $s$ are
both nonzero, one can move diagonally backwards on Table 1,
constructing representations of $Cl(r,s)$ in terms
of the representations of $Cl(r-1,s-1)$, while on the vertical edge of
Table 1, the
representations of $Cl(0,s)$ with $s\ge 2$ can be obtained from
the representations of $Cl(s-2,0)$. These constructions are
trivial, as one just replaces the standard generators with their
2-dimensional representations.
 Hence the nontrivial part is to give
representations of $Cl(r,0)$ in terms of a representation in half dimension.

\vskip 0.2 cm
\noindent
{\bf 3.2 Representations of $Cl(r,s)$ with $s\ge 1$.}
\vskip 0.2 cm

We now give the representations of $Cl(r,s)$ with $s\ge 1$ on $R^{2N}$ in terms
of the representations of $Cl(r-1,s-1)$ or $Cl(s-2,r)$ on $R^N$.

\proclaim Proposition 3.4.  Let $\rho$ be a real
representation of the Clifford algebra $Cl(r,s)$ on $R^N$ and let
$a_i,b_i$
 be the canonical generators of the representation, let $A_i, B_i$  be the
complementary generators satisfying
$$a_i^2+I=0, \quad A_i^2+I=0,\quad b_i^2-I=0,\quad B_i^2-I=0,$$
and let  $J_i$ be the generators of the maximal commuting
subalgebra. Then \item{i.} the canonical generators, the
complementary generators and the generators of the  maximal
commuting subalgebra of the representation of $Cl(r+1,s+1)$ on
$R^{2N}$ are given by
$$\eqalignno{
\tilde{a}_i=& \sigma \otimes a_i,\quad
\tilde{b}_i=  \sigma \otimes b_i,\quad
\tilde{b}_{s+1}=\tau\otimes 1,\quad
\tilde{a}_{r+1}=\epsilon\otimes 1,\cr
\tilde{A}_i=&\sigma \otimes A_i, \quad
\tilde{B}_i=\sigma \otimes B_i,\quad\cr
\tilde{J}_i=&1\otimes J_i,
&(3.1)\cr}$$
\item{ii.}
  the canonical generators,
the complementary generators and
the generators of the  maximal commuting subalgebra of the
representation of $Cl(s,r+2)$ on $R^{2N}$ are given by
$$\eqalignno{
\tilde{a}_i=& \epsilon \otimes b_i,\quad \tilde{b}_i=  \epsilon
\otimes a_i,\quad \tilde{b}_{r+1}=\sigma\otimes 1,\quad
\tilde{b}_{r+2}=\tau\otimes 1,\cr \tilde{A}_i=&\epsilon \otimes
B_i, \quad \tilde{B}_i=\epsilon \otimes A_i,\quad\cr
\tilde{J}_i=&1\otimes J_i. &(3.2)\cr}$$

\vskip 0.2 cm
\noindent
{\bf 3.3 Representations of $Cl(r,0)$.}
\vskip 0.2 cm

We have summarized the structure of the maximal representations of
$Cl(8d+c,0)$  for $c=0,1,3,7$ in Table 2 where it can be seen that
when in passing from a representation of real type  on  $N$
dimensions to a maximal representation in $2N$ dimensions, there
is always a single generator to be added, which can be chosen in
the form $\epsilon\otimes I_N$. On the other hand when the
representation in the half dimension is complex with complex a
structure $J$, we see that one can add $\epsilon\otimes I_N$ and
$\tau\otimes J$, while on quaternionic backgrounds with
quaternionic structures $J_i$, $i=1,2,3$ one can add
$\epsilon\otimes I_N$ and $\tau\otimes J_i$. On the other hand the
existence of the complex or quaternionic structures is related to
the complementary generators with positive squares that can be
added to the representation without increasing the dimension. We
denote the representation in the half dimension as the
``background''.

\vskip 0.3cm \noindent
{\bf Representation of $Cl(8n,0)$ on
$R^{2^{4n}}$:}
We  start with a representation of $Cl(8n,0)$ on
$R^{2^{4n}}$.  The representation is real, hence the maximal
commuting subalgebra is $R$, generated by the identity only.
  $Cl(8n,0)$ can be extended to $Cl(8n,1)$ hence there is a single
complementary generator. The data of the representation is below.
$$\eqalignno{
{\it Canonical\ \  generators}:&\quad
 a_{i}^{(n)},\quad i=1,\dots , 8n,&(3.3a)\cr
 {\it Complementary\ \  generator}:&\quad
\alpha^{(n)}.&(3.3b)}$$

\vskip 0.2   cm \noindent {\bf Representation of $Cl(8n+1,0)$ on
$R^{2^{4n+1}}$:} From the data above  we can obtain the
representations of $Cl(8n+1,0)$ on double dimension. The
representation is complex and extendible to a representation of
$Cl(8n+1,2)$.
$$\eqalignno{
{\it Canonical\ \ generators}:&\quad
b_{i}^{(n)}=\sigma\otimes a_{i}^{(n)},\quad i=1,\dots , 8n,&\cr
& \quad b_{8n+1}^{(n)} =\epsilon\otimes I.&(3.4a)\cr
 {\it Generator \ \ of  \ \ the \ \ commutant:}&\quad
   J^{(n)}=\epsilon\otimes \alpha^{(n)}.&(3.4b)\cr
{\it Complementary\ \  generators}:&\quad
\beta_1^{(n)}=\tau\otimes
I,&\cr
&\quad   \beta_2^{(n)}=\sigma \otimes \alpha^{(n)},&(3.4c)
\cr}$$

\noindent
Note that increasing the number of generators by 1 is trivial, it is sufficient
to tensor the old generators by say $\sigma$ (tensoring with $\tau$ would work
as well) and add the generator $\epsilon \otimes 1$.
The change of the type
from real to complex is tied to the existence of a complementary generator as
follows. A matrix which commutes with
$\epsilon\otimes 1$ has to be either $\epsilon\otimes b$ or $1\otimes c$,
where $b$ is symmetric and $c$ is skew-symmetric. But if $1\otimes c$
commutes with $\sigma\otimes  a_i$, then $a_ic-ca_i=0$, which is not
possible because the background is of real type.
On the other hand
if $\epsilon\otimes b$ commutes with $\sigma\otimes a_i$, then $ba_i+a_ib=0$,
 and as the background admits a complementary generator, it is possible to
choose $J=\epsilon\otimes A_i$.
The existence of one complementary generator, namely $\tau\otimes 1$ is trivial.
The existence of a second one is again tied to the existence of the
complementary generator in the background.

\vskip 0.2     cm
\noindent
{\bf Representation of $Cl(8n+3,0)$ on $R^{2^{4n+2}}$:}
In passing from representations of $Cl(8n+1,0)$ to the representations
of $Cl(8n+3,0)$, the background is of complex type with the
commutant constructed as above.
$$\eqalignno{
 {\it Canonical \ \ generators}:&\quad
c_{i}^{(n)}=\sigma\otimes b_{i}^{(n)},\quad i=1,\dots, 8n+1,&\cr
&\quad c_{8n+2}^{(n)}=\epsilon\otimes I,&\cr
&\quad c_{8n+3}^{(n)}=\tau  \otimes J^{(n)},&(3.5a)\cr
{\it Generators \ \ of \ \  the \   \ commutant:}&\quad
J_1^{(n)} =1\otimes J^{(n)},&\cr
&\quad J_2^{(n)} =\epsilon \otimes \beta_1^{(n)},&\cr
&\quad J_3^{(n)} =\epsilon \otimes \beta_2^{(n)},&(3.5b)\cr
}$$

\noindent As $J$ commutes with the  $b_i$'s we can add the
canonical generator $\tau\otimes J$ which anti-commutes with
$\sigma \otimes b_i$ and $\epsilon\otimes 1$.
 Since the image is quaternionic, we should construct the commutant.
As above, candidates for the commutant are $1\otimes c$ and $\epsilon\otimes
b$. It can be seen that as the background is complex, $c=J$ is possible, hence
the matrices of the quaternionic structure, commuting every canonical generator,
 are $\epsilon\otimes \beta_{8n+1}$.
As the representation is maximal, there are no complementary
generators. \vskip 0.2       cm \noindent {\bf Representation of
$Cl(8n+7,0)$ on $R^{2^{4n+3}}$:} In this case the background is
quaternionic and the dimension of the linear subspace should
increase  by $4$. The $c_i$'s form an anti-commuting set while the
$J_i$'s anti-commute among each
 other but commute with all the $c_i$'s.
Thus in the double dimension we have the representations
$\epsilon\otimes 1$, $\sigma\otimes c_i$ and $\tau \otimes J_i$.  The canonical
generators are thus
$$\eqalignno{
{\it Canonical\  \  generators}:
&\quad
d_{i  }^{(n)}   =\sigma\otimes c_{i}^{(n)},   i=1,   \dots , 8n+3,&\cr
&\quad d_{8n+4}^{(n)}=\epsilon \otimes I, &\cr
&\quad d_{8n+5}^{(n)}=\tau   \otimes   J_1^{(n)},&\cr
&\quad d_{8n+6}^{(n)}=\tau   \otimes   J_2^{(n)},&\cr
&\quad d_{8d+7}^{(n)}=\tau   \otimes   J_3^{(n)}.&(3.6a)\cr
}$$

\noindent
The representation is
real and maximal.
Hence the maximal commuting subalgebra is generated by the identity
only and there are no complementary generators.

\vskip 0.2  cm
\noindent
{\bf Representation of $Cl(8n+8,0)$ on $R^{2^{4n+4}}$:}
Here there is a single generator to be added. As the representation is
real,
the commutant is generated by the identity and there
is a single complementary generator.
$$\eqalignno{{\it Canonical \ \ generators}:
&\quad
a_{i}^{(n+1)}    =\sigma\otimes d_{i}^{(n)}    \quad i=1,\dots 8n+7,&\cr
& \quad a_{8n+8}^{(n+1)}  =\epsilon\otimes I.&(3.7a)\cr
 {\it Complementary \ \  generator}:
&\quad
\alpha^{(n+1)}=\tau\otimes I.&(3.7b)
}$$
These results are summarized in below and in Table 2.

\proclaim Proposition 3.5.
Let $d_i$, $i=1,\dots 8(n-1)+7$ be a set of generators for a
representation of $Cl(8(d-1)+7,0)$ on $R^N$.  Then the set of
canonical generators, the commutant and complementary generators of
representations of $Cl(8d,0)$, $Cl(8d+1,0)$, $Cl(8d+3,0)$ and
$Cl(8d+7,0)$ are determined in terms of these as given in Table 2.

\vskip 0.3cm
\font\big=cmr10 scaled\magstep0
{\big \settabs 5\columns
\hrule
\+&&&&\cr
\+ {\it Clifford Algebra:} & Cl(8d,0) & Cl(8d+1,0) & Cl(8d+3,0) &
Cl(8d+7,0) \cr
\+ {\it Representation space:} & $R^{2^{4d}}$ & $R^{2^{4d+1}}$ & $R^{2^{4d+2}}$ &
                          $R^{2^{4d+3}}$ \cr
\hrule
\+ &&&&\cr
\+ {\it Generators:} &
$a_{i}^{(n)}:$ & $b_{i}^{(n)}:$ & $c_{i}^{(n)}:$ & $d_{i}^{(n)}:$ \cr
\+ & $\sigma \otimes d^{(n - 1)}$ & $\sigma \otimes a_i^{(n)}$ &
$\sigma\otimes  b_i^{(n)}$ & $\sigma\otimes  c_i^{(n)}$ \cr \+ &
$\epsilon \otimes I^{(n - 1)}$ & $\epsilon \otimes I^{(n )}$ &
$\epsilon \otimes I^{(n )}$ & $\epsilon \otimes I^{(n )}$ \cr
 \+ & & & $\tau \otimes J^{(n)}$ & $\tau \otimes J_1^{(n)}$ \cr
 \+ & & &  & $\tau \otimes J_2^{(n)}$ \cr
 \+ & & &  & $\tau \otimes J_3^{(n)}$ \cr \+ &&&&\cr \hrule \+ &&&&\cr
\+ Commutant & - & $J^{(n)}=\epsilon\otimes \alpha^{(n)}$ &
$J_1^{(n)}=1\otimes J^{(n)}$ & \cr \+ & - & - &
$J_2^{(n)}=\epsilon\otimes \beta_1^{(n)}$ & \cr \+ & - & - &
$J_3^{(n)}=\epsilon\otimes \beta_2^{(n)}$ & \cr
\+ &&&&\cr \hrule \+ &&&&\cr
\+ Complementary & $ \alpha^{(n)}=\tau\otimes I^{(n-1)}$ &
$\beta_1^{(n)}=\tau\otimes I^{(n)}$ & & \cr \+ Generators & &
$\beta_2^{(n)}=\sigma\otimes \alpha^{(n)}$ & & \cr \+ &&&&\cr
\hrule }

\vskip 0.3cm
 \noindent
 \font\big=cmr10 scaled\magstep0
{\big Table 2. Construction of the generators, commutant and
 complementary
generators of the representation of
 $Cl(r,0)$ in terms of the data in half dimension.}

\vskip 0.3cm

The product of the generators of a Clifford algebra is called the
``volume element'' and denoted by $\omega$.  The structure of the
volume element in $Cl(r,s)$
is useful in working with iterative constructions.
We first quote below the following result.

\proclaim Proposition 3.6. (Proposition 3.3 in [1]).  Let
$\{e_1,\dots,e_{r+s}\}$ be an orthonormal  set of generators for $Cl(r,s)$ and let
$\omega=e_1\dots e_{r+s}$. Then
$$\omega^2=(-1)^{{n(n+1)\over 2}+s},\eqno(4.4)$$
where $n=r+s$.  Furthermore, for $n$ odd, $\omega$ is a central
element in $Cl(r,s)$ while for $n$ even,
$$
\varphi\omega=\omega\alpha(\varphi),\quad {\rm for \ all}\quad \varphi\in
Cl(r,s),\eqno(4.5)$$
where  $\alpha(\varphi)=\pm \varphi$  respectively for even and odd
elements.

It follows that when $\omega $ is central $\rho(\omega)$ belongs
to the commutant and it can be checked that for $\omega^2=1$,
$\rho(\omega)=I$ while for $\omega^2=-1$, $\rho(\omega)$ is pure
imaginary. Thus for $Cl(8d+3,0)$ and $Cl(8d+7,0)$, as $\omega^2=1$
and $\omega$ is a  central element, hence $\rho(\omega)$  has to
be proportional to identity and the choice of  the sign leads to
inequivalent representations.  In our canonical representation,
the product of the generators of $Cl(3,0)$ is $-1$, and we stick
with this convention. The general form of the volume elements can
be obtained as follows.

\proclaim Proposition 3.7.  Let $a^{(d)}_i$, $b^{(d)}_i$,
$c^{(d)}_i$ and $d^{(d)}_i$ be canonical
generators  as given in Table 2, and assume that
$c^{(0)}_1 c^{(0)}_2 c^{(0)}_3 =-I$. Then for all $d$,
$$
\prod_{i}^{8d}a_i=\tau\otimes 1,\quad
\prod_{i}^{8d+1}b_i=\epsilon\otimes \tau\quad
\prod_{i}^{8d+3}c_i=- I,\quad
\prod_{i}^{8d+7}d_i= I.
$$

\noindent
{\bf Proof.} From Table 2, it is easy to see that
$\prod_{i}^{8d}a_i=(\sigma\otimes 1)(\epsilon\otimes 1)=\tau\otimes
1$.  Then
$\prod_{i}^{8d+1}b_i=(1\otimes \tau)(\epsilon\otimes 1)=\epsilon\otimes
\tau$.  The proofs of the remaining products are similar.
\hfill$\bullet$

\vglue 1cm \noindent
{\bf 4. TRANSFORMATION TO CANONICAL FORMS }
\vskip 0.3cm

 In Section 3 we have given the construction of
canonical forms recursively, starting from lower dimensions. Here
we  consider the converse problem: Given  an orthonormal basis for
the image of $V \subset  Cl(r,s)$ in $R^N$,  find an orthonormal
basis for $R^N$ with respect to which the matrices of the basis
elements have desired canonical forms. From the Wedderburn
theorem, we know that such a basis is unique for Clifford algebras
of real type, while it is determined up to the commutants for
Clifford algebras of complex and quaternionic types.

We would like to start by noting that although the existence  of
the change of basis matrix is guaranteed by the Wedderburn
theorem, its direct determination is not practically feasible. If
an anti-commuting set of matrices with squares $-I$ is given, it
is in principle possible to put them  to canonical forms
iteratively by restricting the change of basis matrix $P$ at each
step, but this procedure gives nonlinear equations for the
components of $P$ and is not useful beyond a set of just  two
matrices.

It is clear that if $Cl(r,s)$ has an irreducible representation on
$R^N$, then any reducible representation on $R^{kN}$ is block
diagonal: One chooses $N$ linearly independent vectors with
respect to which the given matrices have desired forms and then
apply  Gram-Schmidt orthogonalization procedure to obtain a direct
sum splitting and reiterate. Henceforth we consider irreducible
representations only.

We start with the  representations of $Cl(r,s) $ with $s\ge 1$ in Section
4.1.  Section 4.2 is devoted to a detailed the study of the
representations of $Cl(7,0)$ and the general results are given in
Section 4.3.

\vskip 0.3 cm \noindent {\bf 4.1 Representations of $Cl(r,s)$ with
$s\ge 1$.} \vskip 0.2cm

We have  seen that a representation of $Cl(r,s)$ with $s\ge 1$ can be
constructed by tensoring with the representations in half dimension
with $\sigma$ and adding two new generators $\epsilon\otimes 1$ and
$\sigma \otimes 1$.

In the transformation of a given representation
to a canonical one we follow the reverse path.
Assume that we know how to transform a given representation of $Cl(r,s)$ on $R^N$
to a canonical form.  For $Cl(r+1,s+1)$  we
want to identify
one  generator with negative square  as $\epsilon\otimes 1$ and
another  with positive square
 as $\tau\otimes 1$.  Once we find a basis with respect to
which two generators have matrices $\epsilon \otimes 1$ and
$\tau\otimes 1$,  the remaining ones will be automatically of the
form $A_i=\sigma\otimes a_i$, as they should anti-commute with
both $\epsilon \otimes 1$ and $\tau\otimes 1$. The situation is
similar for representations of $Cl(r,s+2)$ where we identify two
generators with positive squares as $\sigma\otimes 1 $ and
$\tau\otimes 1$. We thus start with proving the following Lemma.

 \proclaim Lemma 4.1.   Let $A$, $B$, $C$ be a set of
trace zero anti-commuting linear transformations with
$$A^2+I=0,\quad B^2-I=0,\quad C^2-I=0.\eqno(4.1)$$
Then, there are orthonormal bases of $R^{2N}$ with respect to
which
\item{i.} $B=\sigma\otimes I,\quad C=\tau\otimes I,$
\item{ii.} $A=\epsilon \otimes I,\quad B=\sigma\otimes I,$
\item{iii.} $A=\epsilon\otimes I, \quad B=\tau\otimes I.$

\noindent {\bf Proof.} As $A$, $B$ and $C$ are trace zero, the
$\pm 1$ and $\pm i$ eigenspaces are $N$-dimensional. Thus one can
take $A=\epsilon\otimes I$ or $B=\sigma\otimes I$. We give the
proof of (i) as an example. Let  $X_1,\dots,X_N$ be  an
orthonormal basis for the $+1$ eigenspace of $B$ and define
$Y_i=CX_i$. Computing  $BY_i$ and $CY_i$ it can be seen that $B$
and $C$  have desired canonical forms.  The proofs of (ii) and
(iii) are similar. \hfill$\bullet$

\vskip 0.3cm
\noindent
It follows that given a representation of  $Cl(r,s)$ with $r\ge 1$
 and $s\ge 1$ or for $s\ge 2$, one can
move diagonally  backwards in constructing the representations.

\proclaim Proposition  4.2.  Let
$A_1,\dots,A_{r},B_1,\dots ,B_{s}$
belong to a an irreducible representation of $Cl(r,s)$ on
$R^{2N}$
with
$A_i^2+I=0$,
$B_j^2-I=0$
and assume that the transformation of a given representation to
canonical forms is known on $R^N$.
Then, there   is an orthonormal basis of $R^{2N}$ with respect to
which
\item{i.} if $r\ge 1, s\ge 1$,
$$A_r=\epsilon \otimes I,\quad B_s=\tau\otimes I,\quad
 A_i=\sigma\otimes a_i,\quad i=1,\dots , r-1,\quad
B_j=\sigma\otimes b_j,\quad j=1,\dots ,s-1,\eqno(4.2a)$$
\item{ii.} if $s\ge 2$,
$$B_s=\sigma  \otimes I,\quad B_{s-1}=\tau\otimes I,\quad
 A_i=\epsilon\otimes b_i,\quad i=1,\dots , s-2,\quad
B_j=\epsilon\otimes a_j,\quad j=1,\dots ,r,\eqno(4.2b)$$
\noindent
where
$$a_i^2+I=0,\quad\quad b_j^2-I=0.$$

\noindent
{\bf Proof.}
For (i), using Lemma 4.1 (iii), one can choose an orthonormal basis
$\{X_1,\dots,X_N,Y_1,\dots,Y_N\}$
with respect to which
$A_{r  }$ and $B_{s  }$ have matrices
$\epsilon\otimes I$ and
$\tau\otimes I$.  Then any matrix in the representation which
anti-commutes with these has to be of the form
$\sigma\otimes a$, with   $a^2=- I$ or
$\sigma\otimes b$, with $b^2= I$.
By assumption, the transformations in $N$ dimensions are known, hence
we can put the remaining in canonical forms by orthogonal
transformations on $\{X_1,\dots,X_N\}$.
The proof of (ii) is similar, but there is a twisting as in the proof
of the construction of canonical forms.
\hfill$\bullet$
\vskip 0.3cm

This complete the discussion for the case $s\ge 1$. Although the case
$s=0$ will be discussed in the following section we complete this
section with two remarks on canonical forms for  $Cl(r,0)$.

\vskip 0.2cm \noindent {\bf Remark 4.3.} Lemma 4.1 can be used to
transform representations of $Cl(r+1,0)$ to canonical forms once
the transformation for $Cl(r,0)$ in half dimension is known and a
complementary generator can be identified. This will be the case
for $r=8d$ as Proposition 3.6 implies that the volume element has
square I and is not central. \vskip 0.2cm

\noindent
{\bf Remark 4.4.}
 $Cl(8d+1,0)$ is of complex type and from  Proposition  3.6 it can be
 seen that the product of the generators is a central element with
 square $-I$, hence it is just $J$.
\vskip 0.4cm

\noindent
{\bf 4.2. Representations of $Cl(r,0)$ for $r=3,7$ }

We start by giving  the construction of canonical bases for $Cl(3,0)$
and $Cl(7,0)$.  For $Cl(3,0)$ the generators form a copy of the
quaternionic subalgebra and we obtain the standard generators as in
Proposition 4.5 below.

\vskip 0.4cm \noindent {\bf Representations of $Cl(3,0)$:}

\proclaim Proposition 4.5. Let  $A_1,A_2,A_3$ be an anti-commuting
set of skew-symmetric endomorphisms in $R(4)$ with squares $-I$
and assume that $A_1A_2A_3=-I$. Let $X$ be a unit vector in $R^4$
and define
$$X_1=X,\quad X_2=-A_1X,\quad X_3=-A_2X,\quad X_4=-A_1A_2X.\eqno(4.6)$$
Then $\{X_1,X_2,X_3,X_4\}$ is an orthonormal set with respect to
which the matrices of $A_i$,
$i=1,2,3$ are
$$A_1=\pmatrix{ 0 & 1 & 0 & 0 \cr
               -1 & 0 & 0 & 0 \cr
                0 & 0 & 0 &-1 \cr
                0 & 0 & 1 & 0 \cr}\quad
A_2=\pmatrix  {  0 & 0 & 1 & 0 \cr
                0 & 0 & 0 & 1 \cr
               -1 & 0 & 0 & 0 \cr
                0 &-1 & 0 & 0 \cr   }\quad
A_3=\pmatrix{   0 & 0 & 0 & 1 \cr
                0 & 0 &-1 & 0 \cr
                0 & 1 & 0 & 0 \cr
               -1 & 0 & 0 & 0 \cr   }.\quad
\eqno(4.7)$$

\noindent {\bf Proof.} Using  skew-symmetry and
anti-commutativity, it is easy to check that the  set $\{X, A_1X,
A_2X,$ $ A_1A_2X\}$ is orthonormal. Then by relabelling these
vectors, the matrices of $A_1$ and $A_2$ are of the form
$A_1=\sigma\otimes \epsilon$ and $A_2=\epsilon\otimes 1$, as
above.  The form of $A_3$ follows from the fact that $A_3=A_1A_2$.
\hfill$\bullet$

\vskip 0.4cm \noindent {\bf Representations of $Cl(7,0)$:}

The key fact in the construction above is that the triple product
$A_1A_2A_3$ is proportional to identity. A similar approach does
not work for the representations of $Cl(7,0)$, because the set
consisting of the canonical generators and their products is
linearly independent, consequently, none of the triple products is
proportional to identity. Nevertheless, we show that these triple
products contain an $8$-dimensional abelian subalgebra consisting
of simultaneously diagonalizable matrices with  a common
eigenvector corresponding to the eigenvalue $1$.  Hence, all
triple products are proportional to identity on this one
dimensional subspace and   lead to the desired orthonormal basis.
For this we need to determine conditions under which Clifford
algebra elements commute.

\vskip 0.3cm \proclaim Lemma 4.6.  Let $e_1,\dots,e_{r+s}$ be an
orthonormal set of generators of $Cl(r,s)$ and let
 $\omega$ and $\eta$ be Clifford algebra
elements of orders $a+b$ and $a+c$ of the form
$$\omega=e_{i_1}e_{i_2}\dots e_{i_{a+b}},\quad\quad
 \eta=e_{j_1}e_{j_2}\dots e_{j_{a+c}}.$$
If $\omega $ and $\eta$ have a common factor of order $a$, then
$$\omega\eta=(-1)^{ab+ac+bc}\eta\omega.\eqno(4.8)$$

\noindent
{\bf Proof.} Without loss of generality we can write $\omega=\alpha
\beta $ and $\eta=\alpha \gamma$, where $\alpha$, $\beta$ and $\gamma$
are disjoint. Then the usual rules of exterior algebra applies and we
obtain the result.\hfill$\bullet$
\vskip 0.3cm
\noindent
As an immediate application we can see that a collection of odd
Clifford algebra elements with odd order common factor form an
abelian subalgebra. Hence we have the following.

\vskip 0.4cm \proclaim Proposition 4.7. Let  $A_i$, $i=1,\dots ,7$
be an anti-commuting set of skew-symmetric endomorphisms in $R(8)$
with squares $-I$ and assume that $A_1A_2\dots A_7=I$. Then the
subgroup generated by
$$\mu_1=A_1A_2A_3, \quad \mu_2=A_1A_4A_5,\quad \mu_3=A_2A_4A_6\eqno(4.9)$$
is abelian and have exactly one common eigenvector $X$ with
eigenvalue $+1$.  Hence the set
$$\{A_{123},A_{145},A_{167},A_{246},A_{257},A_{347},A_{356}, I \}\eqno(4.10)$$
where $A_{ijk}=A_iA_jA_k$ is a maximal abelian subalgebra.

\vskip 0.4cm \noindent {\bf Proof. } From Lemma 4.6 it follows
that distinct triple products commute if and only if they have
exactly one common element. Thus the collection given in Eq.(4.10)
is an abelian subalgebra which is clearly maximal in 8-dimensions,
since they generate all  diagonal matrices.
 As this is a commuting set
 of diagonalizable matrices, they can be simultaneously diagonalized.
We have to be careful with the ordering of the factors to
 make sure that they have a common eigenvector with eigenvalue   $1$.
For this we define a commuting set of generators  $\mu_i$ as in
Eq.(4.9). Then, it can be seen that
$$\mu_1\mu_2=A_1A_2A_3A_1A_4A_5=-A_2A_3A_4A_5=-A_1A_6A_7,$$
where the first equality is obtained by anti-commutativity and
using that $A_1^2=-I$, for the second equality uses the fact that
the product is equal to $I$.  Similarly we can obtain
$$\mu_1\mu_3=A_2A_5A_7,\quad \mu_2\mu_3=A_3A_4A_7,\quad
\mu_1\mu_2\mu_3=-A_3A_5A_6.$$

Note that the image of $(\mu_i+I)$ is the $+1$ eigenspace of
 $\mu_i$. Thus if $\mu_1,\mu_2,\mu_3$ had no  common eigenvector with
eigenvalue $1$,
 the product
$$  (\mu_1+I)(\mu_2+I)(\mu_3+I)=I+\mu_1+\mu_2+\mu_3+\mu_1\mu_2+\mu_1\mu_3
+\mu_2\mu_3+\mu_1\mu_2\mu_3\eqno(4.11)$$ would be identically
zero.  From Corollary 3.3, each triple product is trace zero, but
the product in Eq.(4.11) cannot have  trace zero,  hence it is
nonzero. Thus they have a common eigenvector $X$ corresponding to
the eigenvalue $+1$.

Let $\cal A$ be the subalgebra generated by $\{A_1,A_2,A_4\}$. It can be
seen that $\cal A$ acting on $X$ gives a linearly independent set of
vectors in $R^8$ and as $\cal A$ is eight dimensional, $X$ belongs to
a one dimensional subspace.
\hfill$\bullet$

Once we find this preferred direction $X$  on which the 7 triple products
 act as identity,  it is easy to construct the required orthonormal
 basis.

\vskip 0.4cm \noindent {\bf Remark 4.8.} From a computational
point of view, given any basis for a representation, one can use
Gram-Schmidt orthogonalization to obtain an orthonormal
anti-commuting set with squares $-I$ and form the symmetric
matrices $\mu_1$, $\mu_2$ and $\mu_3$ as above.  Then in
8-dimensions, the matrix $(\mu_1+I)(\mu_2+I)(\mu_3+I)$  has rank
$1$, and any of its columns  yield the preferred direction without
any need for eigenvalue computation. This is achieved in OCTAVE
with the command {\it X=orth(Q) }  which gives an orthonormal
basis for the range space of any matrix  $Q$.

We will give in detail the construction of the orthonormal basis.

\vskip 0.4cm \proclaim Proposition 4.9. Let $A_i$, $i=1,\dots ,7$
be an anti-commuting set of skew-symmetric endomorphisms in $R(8)$
with squares $-I$, and let $X$ be a common eigenvector of
$\mu_1=A_1A_2A_3$, $\mu_2=A_1A_4A_5$, $\mu_3=A_2A_4A_6$ with
eigenvalue $1$. Then, with respect to the basis
$$\{X,A_1X,A_2X,\dots ,A_7X\}\eqno(4.12)$$
the linear transformations  $A_i$ have matrices
$$\eqalignno{
A_1=&  - \sigma  \otimes \sigma  \otimes \epsilon\quad
A_2=   -\sigma \otimes \epsilon\otimes     1     \quad
A_3=   - \sigma \otimes\tau \otimes \epsilon     \quad
 A_4=  - \epsilon\otimes 1 \otimes 1 \cr
A_5=&  - \tau    \otimes 1       \otimes \epsilon \quad
 A_6=  - \tau\otimes \epsilon\otimes \sigma       \quad
 A_7=  -  \tau \otimes
\epsilon\otimes \tau  &(4.13)  \cr }$$

\vskip 0.4cm \noindent {\bf Proof. } As the triple products in the
subalgebra generated by the $\mu_i $'s act as identity on   $X$
we can compute the action of all  double products as \settabs
3\columns \+&&\cr  \+ $ A_1A_2X=-A_3X$,    & $A_2A_4X=-A_6X$,  &
$A_3A_7X= A_4X$\cr \+ $ A_1A_3X= A_2X$,    & $A_2A_5X=-A_7X$,  &
$A_4A_5X=-A_1X$\cr \+ $ A_1A_4X=-A_5X$,    & $A_2A_6X= A_4X$,  &
$A_4A_6X=-A_2X$\cr \+ $ A_1A_5X= A_4X$,    & $A_2A_7X= A_5X$,  &
$A_4A_7X=-A_3X$\cr \+ $ A_1A_6X= A_7X$,    & $A_3A_4X=-A_7X$,  &
$A_5A_6X= A_3X$\cr \+ $ A_1A_7X=-A_6X$,    & $A_3A_5X= A_6X$,  &
$A_5A_7X=-A_2X$\cr \+ $ A_2A_3X=-A_1X$,    & $A_3A_6X=-A_5X$,  &
$A_6A_7X= A_1X$.\cr

\vskip 0.4cm Using the above relations we can compute the action
of the $A_i $'s on the basis vectors as given in Table 3 from
which it can be seen that the linear transformation have their
matrices in the desired form with respect to this basis.
\hfill$\bullet$

 \vskip 1cm
\settabs 9\columns
 \+  & & & & & & & & \cr
\+       & $ X$  & $A_1X$& $A_2X$&
$A_3X$& $A_4X$& $A_5X$&$A_6X$& $A_7X$\cr \vskip0.3cm \hrule
\+  & & & & & & & & \cr
\+$A_1$: & $A_1X$& $-X$  &$-A_3X$&$
A_2X$&$-A_5X$&$ A_4X$&$ A_7X$&$-A_6X$\cr
\+$A_2$: & $A_2X$&$
A_3X$&$ -X$&$-A_1X$&$-A_6X$&$-A_7X$&$ A_4X$&$ A_5X$\cr
\+$A_3$: &
$A_3X$&$-A_2X$&$ A_1X$&$   -X$&$-A_7X$&$ A_6X$&$-A_5X$&$ A_4X$\cr
\+$A_4$: & $A_4X$&$ A_5X$&$ A_6X$&$ A_7X$&$
-X$&$-A_1X$&$-A_2X$&$-A_3X$\cr
\+$A_5$: & $A_5X$&$-A_4X$&$
A_7X$&$-A_6X$&$ A_1X$&$   -X$&$ A_3X$&$-A_2X$\cr
\+$A_6$: &
$A_6X$&$-A_7X$&$-A_4X$&$ A_5X$&$ A_2X$&$-A_3X$&$   -X$&$ A_1X$\cr
\+$A_7$: & $A_7X$&$ A_6X$&$-A_5X$&$-A_4X$&$ A_3X$&$
A_2X$&$-A_1X$&$   -X$\cr
\vskip 0.7cm

 \noindent
 \font\big=cmr10 scaled\magstep0
{\big Table 3.
The action of the generators of $Cl(7,0)$ on the basis given by
 Eq.(4.12).}

\vskip 0.3cm
\noindent
{\bf 4.3.  Representations of $Cl(8d+c,0)$, for $d\ge 1$}

Now we give constructions for $Cl(8d+c,0)$ for $d\ge 1$.  The
first step is to get an abelian subalgebra of $Cl(8d+c,0)$.  We
start with determining a maximal abelian subalgebra of $Cl(8d,0)$.

\proclaim Lemma 4.10. Let
$$A^{(1)}_1,\dots ,
A^{(1)}_8,A^{(2)}_1,\dots,A^{(d)}_8 \eqno(4.14)$$ be generators of
$Cl(8d,0)\cong R(2^{8d})$. $Cl(8d,0)$ have an abelian subalgebra
$\cal D$  with $4d$ generators where $\mu^{(i)}_j$, $i=1,\dots
,d$, $j=1,\dots 4$ given by
$$\eqalignno{
\mu^{(1)}_1&=A^{(1)}_1 A^{(1)}_2 A^{(1)}_3,\quad \mu^{(1)}_2
=A^{(1)}_1 A^{(1)}_4 A^{(1)}_5,\quad \mu^{(1)}_3 =A^{(1)}_2
A^{(1)}_4 A^{(1)}_6,&\cr \mu^{(1)}_4& =A^{(1)}_1 A^{(1)}_2
A^{(1)}_3 A^{(1)}_4 A^{(1)}_5 A^{(1)}_6 A^{(1)}_7,&\cr &\dots
\quad \dots&\cr \mu^{(d)}_1&=\omega^{(d-1)} A^{(d)}_1 A^{(d)}_2
A^{(d)}_3,\quad \mu^{(d)}_2 =\omega^{(d-1)} A^{(d)}_1 A^{(d)}_4
A^{(d)}_5,\quad \mu^{(d)}_3 =\omega^{(d-1)} A^{(d)}_2 A^{(d)}_4
A^{(d)}_6,&\cr \mu^{(d)}_4& =\omega^{(d-1)}A^{(d)}_1 A^{(d)}_2
A^{(d)}_3 A^{(d)}_4 A^{(d)}_5 A^{(d)}_6 A^{(d)}_7,&(4.15)\cr }$$
and
$$\omega^{(k)}=A^{(1)}_1\dots  A^{(1)}_8 A^{(2)}_1\dots
A^{(k-1)}_1\dots  A^{(k-1)}_8.\eqno(4.16)$$
$\cal D$ is maximal and has a one dimensional invariant subspace
corresponding to the eigenvalue $+1$.

\noindent {\bf Proof.} The generators given in Eq. 4.15 form an
abelian subalgebra as each of the $\mu^{(i)}_j$'s is and odd
algebra element they have odd common factors. As $\cal D$ is
isomorphic to the subalgebra generated by diagonal matrices in
$R(2^{4d})$ it is maximal.  The proof of the existence of a common
eigenvector and its uniqueness is similar to the proof of
Proposition 4.9. and it is  omitted.

The abelian subalgebras of $Cl(8d+c,0)$ is constructed similarly
but they are not maximal unless $Cl(8d+c,0)$ is of real type and
their invariant subspaces is $2$ or $4$-dimensional according as
they are of complex or quaternionic type.

\proclaim Lemma 4.11. Let
$A^{(1)}_1,\dots ,
A^{(1)}_8,A^{(2)}_1,\dots,A^{(d)}_8, A_1,\dots A_c $
be generators of $Cl(8d+c,0)$ and let $\mu^{(i)}_j$ be as in Lemma
4.10 and let
$\cal D$ be an abelian subalgebra of $Cl(8d+c,0)$.
Then for $c=1,\dots ,3$, $c=4$, $c=5$ and $c=6,7$, $\cal D$ have
respectively $4d$, $4d+1$, $4d+2$ and $4d+3$ generators given by
$$\eqalignno{
c=0,1,2,3:&\quad \mu^{(1)}_1,\dots  ,\mu^{(d)}_4,\cr
c=4      :&\quad\mu^{(1)}_1,\dots  ,\mu^{(d)}_4,
           \quad\mu_{4d+1}=\omega^{(d)}A_1A_2A_3\cr
c=5:      &\quad\mu^{(1)}_1,\dots  ,\mu^{(d)}_4,
           \quad\mu_{4d+1}=\omega^{(d)}A_1A_2A_3,
           \quad\mu_{4d+2}=\omega^{(d)}A_1A_4A_5 \cr
c=6,7:    &\quad\mu^{(1)}_1,\dots  ,\mu^{(d)}_4,
           \quad \mu_{4d+1}=\omega^{(d)}A_1A_2A_3,
           \quad \mu_{4d+2}=\omega^{(d)}A_1A_4A_5,\cr
          & \quad \quad\quad\quad\quad
\mu_{4d+3}=\omega^{(d)}A_2A_4A_6 &(4.17)\cr }        $$
The  invariant subspace of $\cal D$ corresponding to
the eigenvalue $+1$ is one, two or four dimensional  respectively for
$c=0,6,7$,
$c=1,5$ and $c=2,3,4$.

\noindent
{\bf Proof.} The generators in Eq.(4.17) are  commutative as they are
odd algebra elements with odd common factors.
For $c=6,7$, $\cal D$ has $4d+3$ generators in $ R(2^{4d+3})$, hence it
is isomorphic to the diagonal subalgebra and the invariant subspace of
$\cal D$ is one dimensional.  The dimension of other invariant
subspaces
can be determined by similar counting arguments.
\hfill $\bullet$
\vskip 0.3cm

Let $X$ be a unit vector belonging to the invariant subspace of
$\cal D$ corresponding to the eigenvalue $+1$. The  action of
$\cal D$ on $X$ is freely generated by a subset of $4d+j$,
$j=0,1,2,3$ generators. We call these as ``free generators'' and
denote the subalgebra they generate by $\cal A$. We show that the
action of $\cal A$ on $X$ is the required basis.

\vskip 0.3cm \noindent {\bf Remark 4.12.} As the Clifford algebras
$Cl(8d+3,0)$ and $Cl(8d+7,0)$ are isomorphic to a direct sum,
their representations respectively on $R^{2^{4d+2}}$  and
$R^{2^{4d+3}}$  are not faithful and the product of all generators
is identity.  Hence we can  omit the last generator and work with
the representations of $Cl(8d+2,0)$ and $Cl(8d+6,0)$. In
Proposition 4.13 below we do not discuss the cases $c=3$ and
$c=7$. \vskip 0.3cm

We now determine the set of free generators corresponding to $\cal
D$.

\proclaim Proposition 4.13.  Let $\cal D$ be the maximal abelian
subalgebra of the  Clifford algebra $Cl(8d+c,0)$ and let $X$ be a
common eigenvector of $\cal  D$ corresponding to the eigenvalue
$+1$. Then there are $4d$, $4d+1$, $4d+2$ and $4d+3$ free
generators respectively for $c=0$, $c=1$, $c=2$, and $c=4,5,6$,
 given by
$$\eqalignno{
c=0      :&\quad A^{(1)}_1,A^{(1)}_2, A^{(1)}_4,A^{(1)}_8,\dots,
                 A^{(d)}_1,A^{(d)}_2, A^{(d)}_4,A^{(d)}_8,\cr
c=1      :&\quad A^{(1)}_1,A^{(1)}_2, A^{(1)}_4,A^{(1)}_8,\dots,
                 A^{(d)}_1,A^{(d)}_2, A^{(d)}_4,A^{(d)}_8,
                 A_1\cr
c=2      :&\quad A^{(1)}_1,A^{(1)}_2, A^{(1)}_4,A^{(1)}_8,\dots,
                 A^{(d)}_1,A^{(d)}_2, A^{(d)}_4,A^{(d)}_8,
                 A_1,A_2\cr
c=4,5,6  :&\quad A^{(1)}_1,A^{(1)}_2, A^{(1)}_4,A^{(1)}_8,\dots,
                 A^{(d)}_1,A^{(d)}_2, A^{(d)}_4,A^{(d)}_8,
                 A_1,A_2,A_4&(4.18) \cr}$$

\noindent
{\bf Proof.} As the $\mu^{(i)}_k$ acting on $X$ is identity, one of
the $A^{(i)}_j$'s in each of them can be considered as generated by
the other two. Eliminating these we arrive at the set given by
Eq.(4.18) as free
generators.
\hfill$\bullet$

\vskip 0.3cm  We have thus an algorithm for constructing an
orthonormal basis with respect to which a given representation
will have canonical forms.
 The tools developed here can be used for the
problem of transformation a given set of matrices between different
reference frames which may have applications in robotics and computer
graphics.  In the appendix we give  OCTAVE (a Linux shareware package
similar to MATLAB) programs
for the transformations.

\vfill
\eject

\vglue 1cm
\centerline{\bf Appendix A }

OCTAVE programs for the transformation of arbitrary representations to
canonical forms.

\vskip 0.3cm
\noindent
{\bf Canonical forms for a single generator:}

Let $B$ be an endomorphism of $R^{2N}$ with  $B^2-I=0$. Then  $B$
has exactly $N$ eigenvectors corresponding to the eigenvalues $\pm
1$.  Let $X^+=orth(B+I)$ and $X^-=orth(B-I)$.  These are $2N\times
N$ matrices and if $P=[X^+ \ \  X^-]$ then $BP=P\sigma$ and  if
 $P=[X^+ \ \  BX^+]$ then $BP=P\tau$.

Similarly if $A$ is a real
endomorphism of $R^{2N}$ with  $A^2+I=0$, then $A$ has
exactly $N$ eigenvectors corresponding to the eigenvalues $\pm i$.  Let
$X^+=orth(A+iI)$ and $X^-=orth(A-iI)$.  If $P=[X^+\ \  X^-]$ then $AP=iP\sigma$  and  if
 $P={1\over \sqrt{2}}[X^+ +i X^- \ \  -iX^+-X^-]$ then $AP=\epsilon$.

\vskip 0.3cm
\noindent
{\bf Canonical forms for two generators:}

Let $A,B$ be anti-commuting  endomorphism of $R^{2N}$ with
$A^2+I=0$ and  $B^2-I=0$ and let  $X=orth(B+I)$.  If $P=[X \ \
-AX]$ then $BP=P\sigma$  and $AP=P\epsilon$.

Similarly if
 $P={1\over \sqrt{2}}[X+AX\ \ X-AX]$ then $BP=P\tau$  and $AP=P\epsilon$.
In this case as the only matrix anti-commuting with $\epsilon$ and
$\tau$ is $\sigma$, the remaining  endomorphisms  $A_i$ in the
representation are automatically of the form $\sigma\otimes a_i$.

If $B$ and $C$ are anti-commuting endomorphisms with squares $+I$
and $X=orth(B+I)$ then $P=[X\ \ CX]$ results in
$BP=P\sigma$,$CP=P\tau$. It follows that the remaining
endomorphisms  $A_i$ in the representation are automatically of
the form $\epsilon \otimes a_i$.

\vskip 0.3cm
\noindent
{\bf Representation of $Cl(7,0)$:}

Let the representation of the basis elements be $A_1,A_2,\dots
,A_7$. Then the generators of abelian subalgebra are given by
$$
\mu_1= A_1 A_2 A_3,\quad
\mu_2= A_1 A_4 A_5,\quad
\mu_3= A_2 A_4 A_6,\eqno(A.1a)
$$
and $rank((\mu_1+1)(\mu_2+1)(\mu_3+1))=1.$ The common eigenvector
with eigenvalue $+1$ is
$$X=orth((\mu_1+1)(\mu_2+1)(\mu_3+1)).\eqno(A.1b)$$
The action of $Cl(7,0)$ on $X$ is generated by $A_1, A_2,A_4$ and
the transformation matrix is
$$P^{(7)}= \{X \  A_1X\  A_2X\ A_3X\ A_4X\ A_5X\ A_6X\ A_7X\}.\eqno(A.1c)$$
Here the product of the generators is identity and all  triples
products  appear in the
abelian subalgebra.

\vskip 0.3cm \noindent {\bf Representation of $Cl(8,0)$:} Let the
generators of the representation be $A_1,A_2,\dots ,A_7,A_8$. Then
the  abelian subalgebra is generated by $\mu_i$,$i=1,2,3$ and
$$
\mu_4= A_1 A_2 A_3 A_4 A_5 A_6 A_7 \eqno(A.2a)$$
with
$rank ((\mu_1+1)(\mu_2+1)(\mu_3+1)(\mu_4+1)=1.$
The common eigenvector with eigenvalue $+1$ is
$$X=orth((\mu_1+1)(\mu_2+1)(\mu_3+1)(\mu_4+1)).\eqno(A.2b)$$
The action of $Cl(8,0)$ on $X$ is generated by $A_1, A_2,A_4,A_8$
and the transformation matrix is
$$P^{(8)}= \{P^{(7)}\ \  A_8 P^{(7)}\}.\eqno(A.2c)$$

\vskip 0.3cm \noindent {\bf Representation of $Cl(9,0)$:} Let the
generators of the representation be $A_1,A_2,\dots ,A_7,A_8,A_9.$
The abelian subalgebra  is generated by $\mu_i$, $i=1,\dots ,4$ as
before but now $rank((\mu_1+1)(\mu_2+1)(\mu_3+1)(\mu_4+1)=2.$ The
common eigenvectors with eigenvalue $+1$ will belong to the
$2$-dimensional subspace
$$[X_a X_b ]=orth((\mu_1+1)(\mu_2+1)(\mu_3+1)(\mu_4+1)).\eqno(A.3a)$$

If $X$ is any unit vector in the span of $X_a,X_b$, then the
action of $Cl(9,0)$ on $X$ is generated by
 $A_1, A_2,A_4,A_8, A_9$
and the transformation matrix is
$$P^{(9)}= \{P^{(8)}\ \  A_{9} P^{(8)}\}.\eqno(A.3b)$$

\vskip 0.3cm \noindent {\bf Representation of $Cl(11,0)$:} Let the
generators of the representation be $A_1,A_2,\dots
,A_7,A_8,A_9,A_{10},A_{11}.$ The abelian subalgebra are again
generated by $\mu_i$, $i=1,\dots , 4$ and
$rank((\mu_1+1)(\mu_2+1)(\mu_3+1)(\mu_4+1)=4.$ The common
eigenvectors with eigenvalue $+1$ will belong to the 4-dimensional
space
$$[X_a X_b X_c X_d]=orth((\mu_1+1)(\mu_2+1)(\mu_3+1)(\mu_4+1)).\eqno(A.4a)$$

If $X$ is  any unit vector in the span of $X_a,X_b,X_c,X_d$, then
the action of $Cl(11,0)$ on $X$ is generated by $A_1, A_2,A_4,A_8,
A_9,A_{10}$ and the transformation matrix is
$$P^{(11)}= \{P^{(9)} \ \ A_{10} P^{(9)}\}.\eqno(A.4b)$$

\vskip 0.3cm \noindent {\bf Representation of $Cl(15,0)$:} Let the
generators of the representation be $A_1,A_2,\dots ,A_{15}$. Note
that the product of all generators is identity, hence once we put
the first 14 generators in the required form, the last one will
automatically be in the desired format. The maximal abelian
subalgebra is generated by $\mu_i$, $i=1,\dots,4$ as above and
$$\eqalignno{
\mu_5&= A_1 A_2 A_3 A_4 A_5 A_6 A_7 A_8 A_9 A_{10} A_{11}\cr
\mu_6&= A_1 A_2 A_3 A_4 A_5 A_6 A_7 A_8 A_9 A_{12} A_{13}\cr
\mu_7&= A_1 A_2 A_3 A_4 A_5 A_6 A_7 A_8 A_{10} A_{12}
A_{14}&(A.5a)\cr }$$ with
$rank((\mu_1+1)(\mu_2+1)(\mu_3+1)(\mu_4+1)(\mu_5+1)(\mu_6+1)(\mu_7+1)=1$.
The common eigenvectors with eigenvalue $+1$ is
$$X=orth((\mu_1+1)(\mu_2+1)(\mu_3+1)(\mu_4+1)(\mu_5+1)(\mu_6+1)(\mu_7+1).\eqno(A.5b)$$
The action of $Cl(15,0)$ on $X$ is generated by $A_1, A_2,A_4,A_8,
A_9, A_{10},A_{12}$ and the transformation matrix is
$$P^{(15)}= \{P^{(11)}\ \  A_{12} P^{(11)}\}.\eqno(A.5c)$$
\vskip 0.3cm

\vskip 1 cm
 \noindent
{\bf References}

\noindent [1] H.B.  Lawson and  M.L. Michelsohn, {\bf Spin
Geometry}, Princeton U.P., Princeton, NJ, 1989. \vskip 0.2cm
\noindent [2] S. Lang, {\bf  Algebra}, Addison-Wesley, Reading,
MA, 1997. \vskip 0.2cm \noindent [3] Y. Brihaye, P. Maslanka, S.
Giler and P. Kosinski, ``Real representations of Clifford
algebras'',  {\it J. Math. Phys.} vol. 33, (5) (1992), pp.
1579-1581. \vskip 0.2cm \noindent [4] S. Okuba, ``Real
representations of finite Clifford algebras. I. Classification'',
{\it J. Math. Phys.} vol 32, (1991) pp. 1657-1669. \vskip 0.2cm
\noindent [5] L.J. Boya and M. Byrd, ``Clifford periodicity from
finite groups'', J.Phys. A-Math. Gen., vol 32 (18):L201-L205
(1999) \vskip 0.2cm \noindent [6] G. Mullineux, ``Clifford algebra
of three dimensional  geometry'', Robotica, vol 20, pp. 687-697,
(2002). \vskip 0.2cm \noindent [7] O. Roschel, ``Rational motion
design-a survey'', Computer-Aided Design, vol. 30 (3), pp.
169-178, (1998). \vskip 0.2cm \noindent [8] E. Bayro-Corrochano,
``Motor algebra approach for visually guided robotics'', Pattern
Recognition, vol. 35, pp. 279-294, (2002).
 \end